\documentclass[twocolumn,conf]{new-aiaa}
\usepackage[utf8]{inputenc}

\usepackage{graphicx}
\usepackage{amsmath}
\usepackage{siunitx}
\usepackage{longtable,tabularx}
\usepackage[normalem]{ulem} 
\newcommand{\iu}{\mathrm{i}}
\newcommand{\vect}[1]{\boldsymbol{#1}}

\newcommand{\bnabla}{\vect{\nabla}}
\newcommand{\nablaperp}{\nabla_{\!\perp}}

\usepackage{fancyhdr}
\title{Realistic sheared flow profile effects on acoustic impedance eduction in small 3D-ducts}

\author{Lucas~A.~Bonomo\footnote{Postdoctoral Research Fellow, Laboratory of Vibration and Acoustics, lucas.bonomo@lva.ufsc.br, AIAA Member.}, Júlio~A.~Cordioli\footnote{Associate Professor, Department of Mechanical Engineering, julio.cordioli@ufsc.br, AIAA Member.},}
\affil{Federal University of Santa Catarina, 88040-900 Florianópolis, Brazil}

\author{Edward~J.~Brambley\footnote{Reader (Associate Professor), Warwick Mathematics Institute and WMG, E.J.Brambley@warwick.ac.uk, AIAA Senior Member.},}
\affil{University of Warwick, Coventry CV4 7AL, United Kingdom}

\author{Angelo Paduano\footnote{PhD Student, Department of Mechanical and Aerospace Engineering, paduano.angelo@polito.it.} and Francesco~Avallone\footnote{Full Professor, Department of Mechanical and Aerospace Engineering, francesco.avallone@polito.it, AIAA Member.}}
\affil{Politecnico di Torino, 10129, Torino, Italy}

\begin{document}

\onecolumn%
\maketitle

\begin{abstract}%
\noindent
    We investigate the influence of realistic sheared grazing flow on acoustic propagation in three-dimensional rectangular ducts. We show that conclusions reached in the literature about the effects of sheared grazing flow on acoustic propagation in lined ducts are dependent on the flow profiles use in those studies, and that significantly different conclusions are reached once a realistic flow profile is used. We particularly focus on small ducts typical of most experimental impedance eduction facilities, for which velocity gradients are relevant in a significant fraction of the duct cross-section. We assess the effect of simplifying the velocity distribution in the cross-section to either a one-dimensional (2D spanwise-infinite duct) or uniform flow profile.  Three flow profiles are considered, namely (i) the tensorised hyperbolic tangent, (ii) the law-of-the-wall, and (iii) a one obtained from a RANS simulation.  These flow profiles are used as input in numerical simulations, based on the solution of the 3D-Pridmore-–Brown equation, to perform in silico impedance eduction experiments. Results show that realistic flow profiles can be well approximated for acoustic wave propagation in ducts by uniform or 1D flow profiles, provided the bulk Mach number is correctly accounted for, which contrasts with previous findings considering more simplistic flow profiles. The key conclusion of this work is that if viscous effects are negligible and acoustic impedance is a good representation of a lined wall with grazing flow, then the simplification to a uniform flow is a reasonable approximation and traditional eduction methods are sufficiently accurate.
\end{abstract}

\twocolumn 

\section*{Nomenclature}

{\renewcommand\arraystretch{1.0}%
\noindent\begin{tabular}{@{}l @{\quad=\quad} l@{}}

$A^+$ & Van Driest constant \\
$c_0$ & speed of sound, \si{\meter\per\second} \\
$d$ & perforate hole diameter, \si{\meter} \\
$\vect{\hat{\mathrm{e}}_z}$ & the unitary vector in the $z$ direction \\
$f$ & frequency, \si{\hertz} \\
$f_{c,x}$ & cut-off frequency in the $x$-direction, \si{\hertz} \\
$f_{c,y}$ & cut-off frequency in the $y$-direction, \si{\hertz} \\
$H$ & duct height, \si{\meter} \\
$h$ & liner cavity depth, \si{\meter} \\
$\iu$ & imaginary unit, $\sqrt{-1}$ \\
$k_0$ & free-field acoustic wavenumber, \si{\per\meter} \\
$k_s$ & viscous Stokes wavenumber, \si{\per\meter} \\
$k_z$ & axial acoustic wavenumber, \si{\per\meter} \\
$M$ & bulk Mach number \\
$M_{1\mathrm{D}}$ & midspan (1D) average Mach number \\
$p'$ & acoustic pressure, \si{\pascal} \\
$\tilde{p}'$ & complex acoustic pressure amplitude, \si{\pascal} \\
$r$ & normalized radial coordinate \\
$R_{\text{cm}}$ & grazing-flow-induced acoustic resistance \\
$S_r$ & nonlinear resistance slope \\
$S_m$ & nonlinear mass reactance \\
$t$ & perforate facesheet thickness, \si{\meter} \\
$\vect{u}_0$ & mean axial flow profile, \si{\meter\per\second} \\
$U_0$ & mean axial flow velocity, \si{\meter\per\second} \\
$u_\tau$ & friction velocity, \si{\meter\per\second} \\
$u_{\tau,x}$ & friction velocity in the $x$-direction, \si{\meter\per\second} \\
$u_{\tau,y}$ & friction velocity in the $y$-direction, \si{\meter\per\second} \\
$U^+$ & normalized mean velocity \\
$W$ & duct width, \si{\meter} \\
$x,y,z$ & spatial coordinates, \si{\meter} \\
$Z$ & normalized acoustic impedance \\
$Z_{\text{of}}$ & perforate open-face impedance \\
$\theta$ & normalized acoustic resistance \\
$\chi$ & normalized acoustic reactance \\
$\delta$ & boundary-layer thickness, \si{\meter} \\
$\delta^*$ & displacement thickness, \si{\meter} \\
$\delta_t$ & hyperbolic tangent profile shape factor \\
$\kappa$ & von Kármán constant \\
$\mu$ & dynamic viscosity, \si{\pascal\second} \\
$\nu$ & kinematic viscosity, \si{\meter^2\per\second} \\
$\rho_0$ & mean air density, \si{\kilogram\per\meter^3} \\
$\sigma$ & perforate open area ratio \\
$\omega$ & angular frequency, \si{\radian\per\second} \\
$\xi$ & distance from the wall, \si{\meter} \\

\multicolumn{2}{@{}l}{\textbf{Subscripts}} \\

$c$ & centerline value \\
$x,y$ & streamwise-normal and transverse directions \\
$\infty$ & freestream value \\

\multicolumn{2}{@{}l}{\textbf{Superscripts}} \\

$+$ & normalized viscous units \\

\end{tabular}}

\section{Introduction}
    \lettrine{F}{an} noise is a dominant source in modern turbofan aircraft, and its tonal character makes noise reduction critical for aircraft certification. Acoustic liners, typically a honeycomb core between a rigid backplate and a perforated facesheet, are the main passive treatment. Explicitly modeling the physics of a whole liner remains too expensive due to multiscale interactions --- such as acoustic-induced flow and boundary layer --- although small-scale high-fidelity simulations have recently become feasible \citep{zhang2016Numerical, paduano_impact_2025}. Instead, liners are described by their acoustic impedance, $\Tilde{Z}(\omega) = \theta + \iu\chi$, with $\theta$ the resistance and $\chi$ the reactance. This frequency-dependent property serves as a boundary condition in acoustic models of aeroengine nacelles.
    
    The impedance of an acoustic liner depends on liner geometry~\citep{guess1975Calculation, jones2002effects} and operating conditions such as the Sound Pressure Level (SPL)~\citep{murray2012development}, grazing flow velocity amplitude~\citep{guess1975Calculation}, and flow profile~\citep{kooi1981Experimental}. Today, impedance eduction is the standard experimental approach for characterizing liners with grazing flow. It infers impedance from in-duct acoustic measurements and a duct propagation model, offering simpler instrumentation, higher repeatability, and broad applicability across liner designs when compared to the earlier in situ (Dean’s) method.~\citep{watson1999validation, jing2008straightforward, elnady2009validation, ferrante2016back, bonomoComparisonSituImpedance2024, rashidi20253D, troian2017Broadband, zhao2025DeepLearningBased}.
    
    Impedance eduction methods rely on acoustic propagation models for ducts. Common simplifications are that (i) the flow is uniform, and (ii) the Ingard--Myers Boundary Condition (IMBC)~\citep{ingard1959influence,myers1980acoustic} can be used to take into account the thin boundary layer effects on acoustic propagation. These simplifications are currently under scrutiny in the scientific community. In principle, the acoustic impedance of liners should remain independent of the incident acoustic field, under the assumption of local reactivity. However, experiments have shown different impedance values when the wave propagation direction changes between upstream- and downstream-propagating waves, suggesting flaws in the modeling used for impedance eduction~\citep{renou2011Failure, boden2016Effect, roncen2019influence, boden2017Comparison}. Traditionally, the upstream/downstream discrepancy has been reported to manifest as a scissor-like behavior of the educed resistance, in which the upstream source tends to yield lower resistance values at low frequencies, and higher resistance values at higher frequencies, than the downstream source, i.e., a crossing of the upstream- and downstream-educed resistance curves as a function of frequency. The crossing point between the two resistance curves is usually observed near the liner resonance
    
    Several attempts have been made to include governing equations, capturing actual liner physics, to solve for this discrepancy. For instance, Ref.~\cite{weng2018Flow} found that considering the effects of the viscosity in the acoustic propagation is unable to account for the discrepancy between upstream and downstream and collapse the educed impedances. Ref.~\cite{nark2018assessment} found that small changes to the mean flow velocity were able to reduce the gap between the upstream and downstream educed impedances, but not collapse the curves. Ref.~\cite{roncen2020WavenumberBased} showed that the difference persists even when considering a three-dimensional inviscid sheared flow profile and a no-slip mean flow boundary condition at the lined wall, suggesting that liner physics is not fully captured. Novel boundary conditions have been introduced to better describe liner physics, but none have accurately predicted acoustic wavenumbers when the experimental configuration is changed~\citep{spillere2020Experimentally}.
    
    Recently, Ref.~\citep{bonomo2026Effect} extended the work of \citet{nayfeh1974Effect} by examining how velocity profile formulations affect impedance eduction. Using the Pridmore–-Brown Equation (PBE)~\cite{pridmorebrown1958sound}, wavenumbers for realistic profiles were compared against those obtained with the uniform flow assumption and the IMBC. The results showed that the IMBC~\cite{ingard1959influence,myers1980acoustic} is adequate for small ducts and low Mach numbers when using realistic profiles, but errors grow when using simplified profiles, increasing with duct width and mean flow velocity. A key limitation of Ref.~\citep{bonomo2026Effect} was the restriction to two-dimensional ducts (i.e.\ 1D flow profiles), leaving the role of three-dimensional effects unresolved. These observations are consistent with the findings of Ref.~\mbox{\citep{yangShearFlowEffects2024}}, who used a power-law velocity profile and showed that shear flow has a limited impact on the educed impedance at low Helmholtz numbers (i.e., at low frequencies and in small ducts). Since the choice of velocity profile formulation strongly impacts the educed impedance, the findings of Ref.~\mbox{\cite{roncen2020WavenumberBased}} may be unreliable, and a reassessment with more realistic flow representations is necessary to properly understand sheared-flow effects in 3D ducts.
    
    For clarity, we distinguish between the dimensionality of the duct, which defines the physical geometry, and that of the flow profile or solver, which determines the mathematical formulation. A three-dimensional (3D) duct refers to a rectangular geometry supporting variations in both the spanwise and wall-normal directions, for which the mean flow profile is two-dimensional. Conversely, a two-dimensional (2D) duct restricts the problem to a single wall-normal velocity profile, resulting in a one-dimensional (1D) formulation of the PBE. Throughout this paper, we refer to the 2D PBE as the partial differential form governing acoustic propagation in 3D ducts, and to the 1D Pridmore–Brown equation as the simplified ordinary differential form used in 2D ducts.
    
    In this paper, we reassess the impact of simplifying the three-dimensional nature of acoustic wave propagation in lined rectangular ducts to either a two-dimensional domain or a uniform flow case, under conditions typical of impedance eduction facilities. Building on our recent findings on the role of axial velocity profile shape~\citep{bonomo2026Effect}, we revisit earlier conclusions, including those of Ref.~\citep{roncen2020WavenumberBased}, which relied on simplified flow assumptions. In this paper, we perform virtual \emph{in silico} experiments using the full 2D version of the PBE as the reference model; using this, we compute the least-attenuated mode wavenumbers for different velocity profiles and then apply the straightforward impedance eduction method to quantify how uniform flow or 2D approximations may distort experimental outcomes. Importantly, the present work demonstrates that the conclusions drawn by Ref.~\mbox{\citep{roncen2020WavenumberBased}} regarding the role of sheared flow in three-dimensional ducts warrant reconsideration, as they were obtained using simplified velocity profiles that may not adequately represent a realistic boundary-layer effects. In particular, special attention is given to the role of the effective mean flow Mach number used in the eduction process. As will be shown, ensuring consistency between the cross-section averaged Mach number of the reference configuration and that assumed in the eduction model plays a key role on upstream/downstream discrepancies, and may outweigh the influence of the specific shear profile representation.
    
    This paper is organized as follows. Section~\ref{sec:equations} presents the governing equations for 3D duct acoustics with sheared grazing flow. Section~\ref{sec:materials} describes the setup for the numerical experiments, and the main results regarding the flow profile shape effects on impedance eduction in 3D ducts are presented in Section~\ref{sec:results}. Finally, the concluding remarks are presented in Section~\ref{sec:conclu}.    


\section{Governing equations} \label{sec:equations}

\begin{figure}[htb!]
        \centering
        \includegraphics[width = 3.15in]{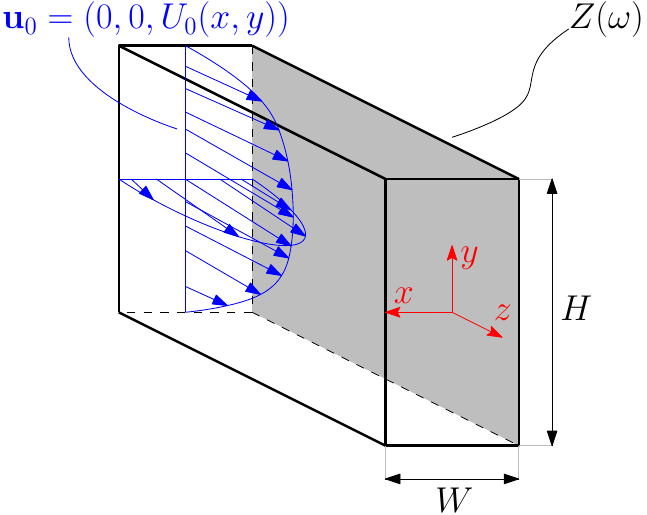}
        \caption{Schematic duct and coordinates system adopted in this work.}
        \label{fig:schematicDuct}
    \end{figure}

    For the purpose of this study, the infinite rectangular duct depicted in Fig.~\ref{fig:schematicDuct} is considered. The duct cross-section has width $W$ and height $H$. The axial flow has flow profile $\vect{u}_0 = U_0(x,y)\vect{\hat{\mathrm{e}}_z}$, where $\vect{\hat{\mathrm{e}}_z}$ is the unitary vector in the $z$ direction. Neither the flow profile nor the impedance vary in the axial direction. This implies that the flow is fully developed in the duct. The wall located at $x = -W/2$ has a locally-reacting frequency-dependent impedance, $Z(\omega)$, while all the other walls are acoustically rigid. 
    
    For the purpose of this work, we will assume that the in-duct acoustic propagation can be described by the PBE \citep{pridmorebrown1958sound}, such that
    \begin{multline}
        (\iu \omega + \vect{u_0\cdot\nabla}) \left( \dfrac{1}{c_0^2} (\iu \omega + \vect{u_0 \cdot \nabla})^2 \tilde{p}' - \nabla^2 \tilde{p}' \right) \\+ 2 \dfrac{\partial}{\partial z} \big(\bnabla \tilde{p}' \vect{\cdot \,\nabla}\! U_0\big) = 0, \label{eq:PB_generalForm}
    \end{multline}
    where $\tilde{p}'$ is the acoustic pressure with assumed monochromatic time dependence $\exp ( \iu\omega t )$, $c_0$ is the speed of sound, $\iu = \sqrt{-1}$ is the complex imaginary unity and $\bnabla = (\partial/\partial x, \partial/\partial y, \partial/\partial z)$. Considering the axial invariance of the problem, we assume an axial dependence of the form $\tilde{p}'(x,y,z) = \tilde{p}'(x,y) \exp\{-\iu k_z z\} $, where $k_z$ is the axial wavenumber, so Eq.~\eqref{eq:PB_generalForm} can be written as
    \begin{multline}
        \left(\nablaperp^2 + \dfrac{\omega^2}{c_0^2}\right)\tilde{p}' - k_z \left( \dfrac{U_0}{\omega}  \nablaperp^2 - \dfrac{2}{\omega} \nablaperp U_0 \cdot \nablaperp + \dfrac{3 \omega U_0}{c_0^2} \right) \tilde{p}' \\ - k_z^2 \left( 1 - \dfrac{3U_0^2}{c_0^2} \right) \tilde{p}' - k_z^3 \left[ \dfrac{U_0}{\omega} \left( \dfrac{U_0^2}{c_0^2} - 1\right) \right] \tilde{p}' = 0, \label{eq:PBE_kz}
    \end{multline}
    where $\vect{\nablaperp} =  (\partial/\partial x, \partial/\partial y, 0)$.        
    
    The acoustic velocity $\vect{u'}$ is then given in terms of the acoustic pressure by
    \begin{equation}
    \vect{u'} = \frac{\iu}{\rho_0(\omega-U_0k_z)}\bnabla\tilde{p}' - \frac{\bnabla\tilde{p}'\vect{\cdot\,\nablaperp}\!U_0}{\rho_0(\omega-U_0k_z)^2}\vect{\hat{e}_z}.
    \end{equation}%

    As boundary conditions, at rigid walls, the normal acoustic velocity $\vect{u}'$ vanishes, such that
    \begin{equation}
        \vect{u}'\cdot\vect{\hat{\mathrm{n}}} = 0, 
    \end{equation}
    where $\vect{\hat{\mathrm{n}}}$ is a unitary normal vector pointing into the wall. Non-slip boundary condition is assumed, therefore the locally reacting impedance boundary condition can be written as
    \begin{equation}
        Z = \dfrac{1}{\rho_0 c_0} \dfrac{\tilde{p}'}{\vect{u'\cdot\hat{\mathrm{n}}}}, \label{eq:Z}
    \end{equation}
    where the air characteristic impedance $\rho_0 c_0$ is used as a normalization factor and $\rho_0$ is the air density.

    \subsection{Uniform flow case}

        If a uniform flow is assumed, i.e., $U_0 (x,y) \equiv M c_0$, where $M$ is the bulk (cross-section average) Mach number, the PBE~\eqref{eq:PB_generalForm} reduces to the Convected Helmholtz Equation (CHE),
        \begin{equation}
         \nabla^2 \tilde{p}' + \left(k_0 -\iu M \frac{\partial}{\partial z}\right)^2 \!\!\tilde{p}' = 0. \label{eq:CHE}
        \end{equation}
        where $k_0 \equiv \omega /c_0$ is the free-field wavenumber.
    
        The impedance boundary condition, given by Eq.~\eqref{eq:Z}, is valid provided the mean flow velocity vanishes at the wall. With a uniform flow model, such as the CHE, the flow velocity adjacent to the wall is non-zero, although in reality there is a boundary layer at the wall.  The refractive effects that occur within this boundary layer must be taken into account by the boundary condition. The most common approach is the IMBC \citep{ingard1959influence, myers1980acoustic}, which can be written as
        \begin{equation}
            Z_{\mathrm{eff}, \mathrm{IMBC}} = \frac{1}{\rho_0c_0}\dfrac{\tilde{p}'}{\vect{\vect{u}'} \cdot \vect{\hat{\mathrm{n}}}} = \dfrac{\omega Z}{\omega - c_0M k_z}. \label{eq:k_dep_IMBC}
        \end{equation}
        It is emphasized that a uniform mean flow in a straight duct is assumed in the present formulation. Under these conditions, curvature-related terms associated with the Myers boundary condition do not contribute, and no boundary-layer effects are modeled.

    \subsection{Eigenvalue problem} \label{sec:solver}

        We seek to describe the governing equations as a generalized eigenvalue problem. One can rewrite the PBE~(Eq.~\eqref{eq:PBE_kz}) as
        \begin{equation}
            (\vect{\mathrm{A}}_0 + \vect{\mathrm{A}}_1 k_z + \vect{\mathrm{A}}_2 k_z^2 + \vect{\mathrm{A}}_3 k_z^3) \vect{\tilde{p}}' = \vect{0}, \label{eq:genEig}
        \end{equation}
        where the $\mathrm{A}_j$ terms involve differentiation in $x$ and $y$ and multiplication by the frequency $\omega$, by the mean flow $U_0$ and by its $x$- and $y$-derivatives, and $\vect{\tilde{p}}'$ is the discretized acoustic pressure.  In the present work, we follow a pseudo-spectral strategy similar to \citet{boyer2011Theoretical}, where the problem is discretized by projecting it onto a Gauss--Lobatto grid using Chebyshev polynomials as basis.  Finally, to solve the cubic generalized eigenvalue problem given by Eq.~\eqref{eq:genEig}, auxiliary variables are introduced to linearize the eigenvalue problem. Specifically, variables of the form $\vect{\tilde{p}}_p = k_z \vect{\tilde{p}}_{p-1}$ for $p>0$ are defined following the standard approach described in \cite[p.~129]{boyd2001Chebyshev}. This reformulation yields an equivalent sparse linear eigenvalue problem, which is solved using the \texttt{eigs} function in \textsc{Matlab}.
    
        In order to apply a lined wall boundary condition to the generalized eigenvalue problem, we rewrite Eq.~\eqref{eq:Z} as
        \begin{equation}
            \dfrac{\partial \tilde{p}'}{\partial x} n_x + \dfrac{\iu \omega}{c_0 Z}\tilde{p}' = 0,
        \end{equation}
        where $n_x = -1$ at $x = -W/2$. For the hard wall opposite to the liner, the corresponding boundary condition is
        \begin{equation}
            \dfrac{\partial \tilde{p}'}{\partial x} = 0.
        \end{equation}
        Similarly, for the hard walls at $y = \pm H/2$, one may write
        \begin{equation}
            \dfrac{\partial \tilde{p}'}{\partial y} = 0.
        \end{equation}

        In this work, we employed a collocation grid with $N_x \times N_y = 151 \times 41$ points in the $x$ and $y$ directions, respectively. The convergence analysis was carried out independently in each direction following the same procedure as Ref.~\cite{bonomo2026Effect}, yielding an estimated numerical error of $\varepsilon < \num{1e-6}$. Note that $N_y < N_x$ despite $H > W$, owing to the presence of only acoustically hard walls in the $y$ direction.

        It is also worth noting that early finite-element discretizations of the Pridmore--Brown equation revealed the presence of additional continuous spectra associated with convected vorticity, often referred to as hydrodynamic modes \mbox{\cite{brambley2009fundamental}}. While these modes are not explicitly analyzed in the present work, they form part of the broader theoretical framework of duct acoustics with mean flow.


\section{\textit{In Silico} Experiments} \label{sec:materials}

    In this section, we explore the effects of considering a 2D flow profile while solving for the acoustic propagation in a duct. We propose to revisit the work of \citet{roncen2020WavenumberBased} in light of the influence of the flow profile shape formulation, by means of a numerical experiment. In this work, we consider a fully developed flow in a 3D duct.

    \subsection{Mean flow profile}
    
\begin{figure*}
        \centering
        \includegraphics[width=\linewidth]{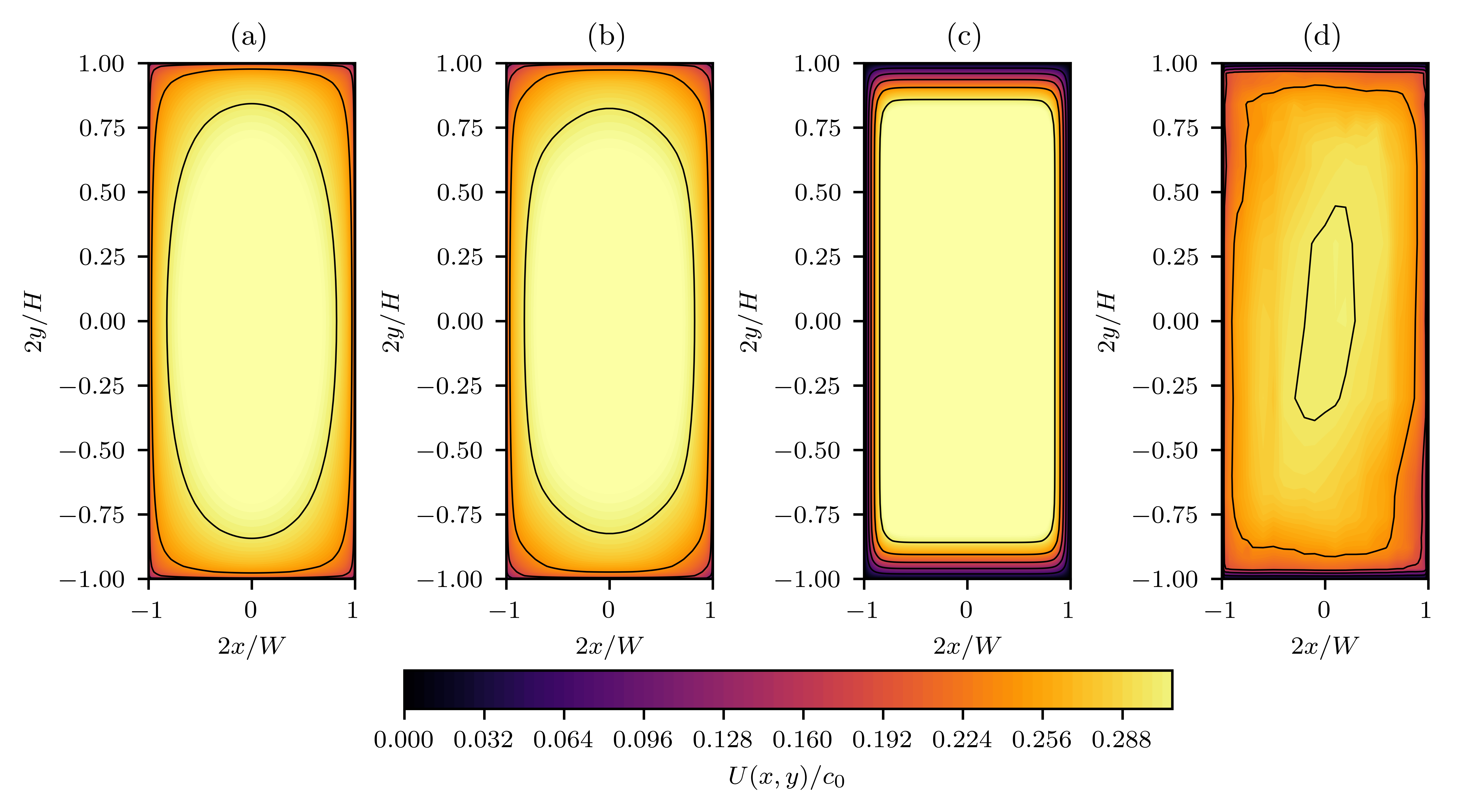} \\
        \includegraphics{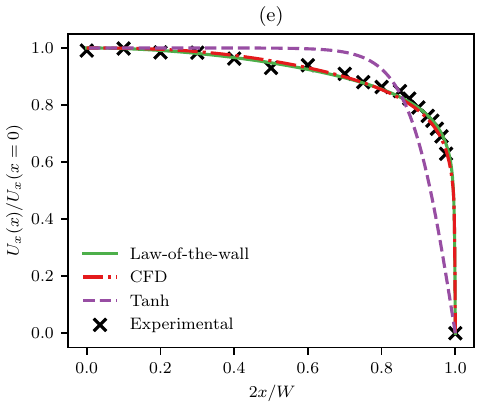}
        \caption{Flow profiles considered in this work: (a) Tensorized law of the wall; (b) CFD evaluated, and; (c) tensorized hyperbolic tangent. (d) Displays the  experimental data measured in the LITR/UFSC for a slightly smaller average velocity, for the purposes of comparison. Isolines represent contour axial velocity. (e) shows the comparison of the different profiles for \mbox{$y = 0$} and \mbox{$x\in[0,W]$}, normalized by the maximum velocity.}
        \label{fig:flowProfiles}
    \end{figure*}

    The definition of the velocity profile for turbulent flows in a rectangular duct is a non-trivial task. The presence of secondary flows leads to the transport of flow momentum from the center of the duct to the corners, which is then pushed back toward the center along the walls to preserve continuity~\cite{nakayama1983Calculation}. This creates complex flow patterns that make obtaining an explicit formulation for the 2D velocity profile challenging. To the best of the authors' knowledge, no such formulation is currently available in the literature. A common simplification is to adopt a tensorization of 1D flow profiles in the $x$ and $y$ directions to construct a 2D flow profile~\citep{roncen2020WavenumberBased}. Here, “tensorized” refers to the construction of a two-dimensional velocity field by combining independent one-dimensional profiles defined along each transverse direction. In this work, we consider three flow profiles for the rectangular duct: two obtained through the traditional tensorization procedure and one obtained with the assistance of Computational Fluid Dynamics (CFD). The three flow profiles considered in this work are shown in Fig.~\ref{fig:flowProfiles}, where an experimental mean flow profile measured in the Liner Impedance Test Rig at the Federal University of Santa Catarina (LITR/UFSC) for a bulk Mach number of 0.278 is also included for comparison.

        We use the dimensions of the LITR/UFSC of $W = \SI{40}{\milli\meter}$ and height $H = \SI{100}{\milli\meter}$, as representative of a typical liner testing facility. The frequency range is defined from 500 to $3000\,\mathrm{Hz}$, with steps of $100\,\mathrm{Hz}$. We consider a bulk Mach number of $M = 0.3$. For reference, the cut-off frequencies of the rigid duct are determined by the duct cross-sectional dimensions. In the streamwise-normal ($x$) direction, the first cut-off frequency is given by $f_{c,x} = c_0/(2W)$, yielding $f_{c,x} \approx \mbox{\SI{4.29}{kHz}}$. Similarly, in the transverse ($y$) direction, the first cut-off frequency is $f_{c,y} = c_0/(2H)$, resulting in $f_{c,y} \approx \mbox{\SI{1.72}{kHz}}$.
        
        For the tensorised flow profiles, we need formulations for 1D duct flow (channel flow). In this work, two formulations are considered. A commonly employed formulation is the hyperbolic tangent profile introduced by \citet{rienstra2008spatial}, that was for instance used by \citet{roncen2020WavenumberBased} to study 2D flow profile effects on impedance eduction. The hyperbolic tangent profile is given by
        \begin{align}\label{eq:tanh}
            &M(r) = M_c \left[ \tanh{\left(\dfrac{1-r}{\delta_t}\right)} \right.\hfill\\\notag&\left. + \big(1-\tanh{(1/\delta_t)\big)\!\left( \dfrac{1+\tanh{(1/\delta_t)}}{\delta_t}r +(1+r)\right)}(1-r)\right]\!, 
        \end{align}
        where $M_c$ is the centerline Mach number, $r$ is the radial position and $\delta_t$ is a shape factor. In this work, we use the transformations of coordinates $r(x) = |W/2 - x|$, $r(y) = |H/2 - y|$ in order to obtain the flow profiles in the $x$ and $y$ directions, respectively. 
    
        We also want to consider a more realistic representation of the boundary layer velocities profile. For this purpose, a universal law-of-the-wall is used. We consider the formulation given by \citet{van1956turbulent}, such that
        \begin{equation}
            U^+ =  \int_0^{y^+}\!\!\!\! \dfrac{2}{1 + \sqrt{1 + 4 \kappa^2 {y^+}^2(1-\exp(-y^+/A^+)})^2} \mathrm{d}y^+ + \Pi, \label{eq:vanDriest}
        \end{equation}
        where $U^+ \equiv U_0 / u_\tau$ is the flow profile normalized by the friction velocity $u_\tau$, $\kappa \approx 0.42$ is the von Kármán constant, $A^+ \approx 27$ is the Van Driest constant, and $y^+ = \xi u_\tau / \nu$ is the distance from the wall, $\xi$, normalized to viscous lengths, with $\nu$ being the air kinematic viscosity. As will be discussed later, for the small ducts considered in this study, the boundary layer can extend to the entire half-duct width. To ensure that the derivative of the profile is continuous at the duct centreline, we propose adding a quadratic term to Eq.~\eqref{eq:vanDriest}, denoted by $\Pi$, which is given by
        \begin{equation}
             \Pi = \dfrac{2(y^+_{\text{max}} - y^+)}{1 + \sqrt{1 + 4 \kappa^2 {y^+_{\text{max}}}^2(1-\exp(-y^+_{\text{max}}/A^+)})^2} \left(\dfrac{y^+}{y^+_{\text{max}}}\right)^2\!,\label{eq:Pi}
        \end{equation}
        where $y^+_{\text{max},x} = W u_{\tau,x} / \nu / 2$ and $y^+_{\text{max},y} = H u_{\tau,y} / \nu / 2$ are the distances from the wall to the centreline in viscous lengths, for the $x$ and $y$ directions, respectively. In this work, the friction velocities $u_{\tau,x}$ and $u_{\tau,y}$ were obtained by fitting Eq.~\mbox{\eqref{eq:vanDriest}} to the experimental mean flow data measured at the LITR and shown in Fig.~\mbox{\ref{fig:flowProfiles}}d. To estimate $u_{\tau,x}$, experimental data along the line $0 \leq x \leq W/2$ at $y=0$ were used, whereas for $u_{\tau,y}$ the fitting was performed using data along $0 \leq y \leq H/2$ at $x=0$. This procedure yielded $u_{\tau,x} = \SI{3.95}{\meter\per\second}$ and $u_{\tau,y} = \SI{3.71}{\meter\per\second}$. This procedure is similar to the one employed in Ref.~\mbox{\cite{bonomo2026Effect}} for fitting one-dimensional mean-flow profiles to experimental data.

        The formulations for the hyperbolic tangent and the turbulent wall law are used to generate the midspan flow profiles in the $x$ and $y$ directions ($y=0$ and $x=0$, respectively). The shape factor $\delta_t$ of the hyperbolic tangent velocity profile was adjusted independently in the streamwise ($x$) and transverse ($y$) directions so that the resulting boundary-layer thickness matches that obtained from the corresponding law-of-the-wall profiles employed in this study. Based on this criterion, the boundary-layer displacement thicknesses of the law-of-the-wall profiles are $\delta_x^* = 1.70~\mathrm{mm}$ and $\delta_y^* = 3.93~\mathrm{mm}$, which lead to shape factors $\delta_t = 12.27\%$ and $\delta_t = 11.34\%$ in the $x$- and $y$-directions, respectively. This choice ensures a consistent comparison between the different mean-flow models and follows the methodology proposed in our recent work Ref.~\mbox{\cite{bonomo2026Effect}}. These two 1D flow profiles are tensorised to a 2D flow profile and rescaled for the bulk Mach number by
        \begin{equation}
            \dfrac{U_0(x,y)}{c_0} = M \dfrac{U_x(x)}{\langle U_x(x)\rangle} \dfrac{U_y(y)}{\langle U_y(y)\rangle},
        \end{equation}
        where the operator $\langle \cdot \rangle$ denotes averaging over the function domain. The bulk Mach number is the spatial 2D-average of the velocities profile, therefore
        \begin{equation}
            M = \dfrac{1}{WH} \int_{-W/2}^{W/2}\int_{-H/2}^{H/2} \dfrac{U_0(x,y)}{c_0} \mathrm{d}y \mathrm{d}x.
        \end{equation}
        Another important quantity is the midspan average Mach number, which is often considered by research groups in in-duct propagation models~\citep{roncen2020WavenumberBased, jingInvestigationStraightforwardImpedance2015}, even though it neglects the inherent two-dimensional nature of the velocities profile, and is given by
        \begin{equation}
            M_{1\mathrm{D}} = \dfrac{1}{W}\int_{-W/2}^{W/2} \dfrac{U_0(x,y = 0)}{c_0}  \mathrm{d}x.
        \end{equation}
        Although \citet{roncen2020WavenumberBased} investigated the impact of considering either $M$ or $M_{1\mathrm{D}}$ in the wavenumbers calculations, no report is given on the impact of this choice in impedance eduction, which we will explore in this study.

        The numerical simulations of the mean flow within the duct were conducted using the commercial CFD software {STAR-CCM+}, employing a Reynolds-Averaged Navier–Stokes (RANS) framework to resolve the time-averaged flow field. The turbulence closure was achieved using the Shear Stress Transport (SST) $k-\omega$ model developed by \citet{menter2003ten}, which offers improved accuracy in capturing near-wall behavior.
        The simulations are carried out in a rectangular computational domain of dimensions \( L_x \times L_y \times L_z = 2.5h \times h \times 300h \), where \( h = H/2 = \SI{20}{\milli\meter} \) denotes the half-height of the channel. The coordinate directions \( x \), \( y \), and \( z \) correspond to the spanwise, wall-normal, and streamwise axes, respectively (Fig.~\ref{fig:schematicDuct}). To reduce computational costs while retaining the essential flow physics, only a quarter-section of the duct is simulated. Reflection boundary conditions are imposed along the lateral and vertical walls to account for geometric symmetry and emulate the influence of the complete duct configuration.  

        The governing equations are discretized on a non-uniform grid consisting of \( N_x \times N_y \times N_z = 200 \times 80 \times 250 \) cells. The streamwise discretization features anisotropic refinement, designed to capture the self-similar characteristics of the fully developed flow region. In the spanwise and wall-normal directions, grid points are clustered near the wall using a hyperbolic stretching function that ensures smooth cell-size transitions and coarsening towards the freestream. This configuration yields wall-normal spacings with \( \Delta y^+_{\text{min}} \approx 0.58 \) at the wall, and spanwise spacings between \( \Delta x^+_{\text{min}} \approx 0.56 \) and \( \Delta x^+_{\text{max}} \approx 1.4 \).  
        
        At the inflow, a prescribed velocity boundary condition is applied, while a pressure outlet is set at the downstream boundary. The freestream Mach number is \( M_{\infty} = 0.3 \), and the friction Reynolds number is \( \mathrm{Re}_{\tau} = (\delta u_{\tau}) / \nu \approx 5500 \).        

        \subsection{Test Liner}
        
\begin{figure}[htb!]
            \centering
            \includegraphics[width=\linewidth]{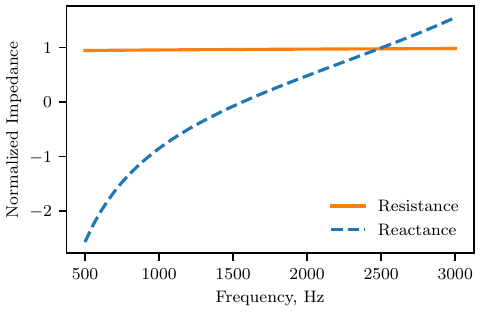}
            \caption{Reference impedance considered in the numerical experiment.}
            \label{fig:refImpedance}
        \end{figure}

        The impedance predicted by the Goodrich impedance model \citep{yu2008validation} for a typical single-degree-of-freedom liner sample is used (for details, see \hyperlink{ap:utas}{the Appendix A}). The liner geometry corresponds to the 3D-printed liner sample which has been extensively used by the authors in previous (and ongoing) experimental and high-fidelity simulation investigations \citep{quintino2025Comparison, pereira2023Validation}. The perforate facesheet has holes with diameter of \SI{1.17}{\milli\meter}, thickness of \SI{0.55}{\milli\meter} and a percentage of open area of approximately \SI{8.8}{\percent}, with a cavity depth of \SI{38.1}{\milli\meter}. The reference impedance can be seen in Fig.~\ref{fig:refImpedance}.
        Within this framework, the present work serves as a baseline study, providing a controlled reference against which more general formulations accounting for spatially varying and nonlinear impedance effects may be assessed in future work. 


\section{Results and Discussion} \label{sec:results}

    \subsection{Wavenumbers of the least-attenuated mode}
    
\begin{figure*}[t]
            \centering
            \includegraphics[width = 6.4in]{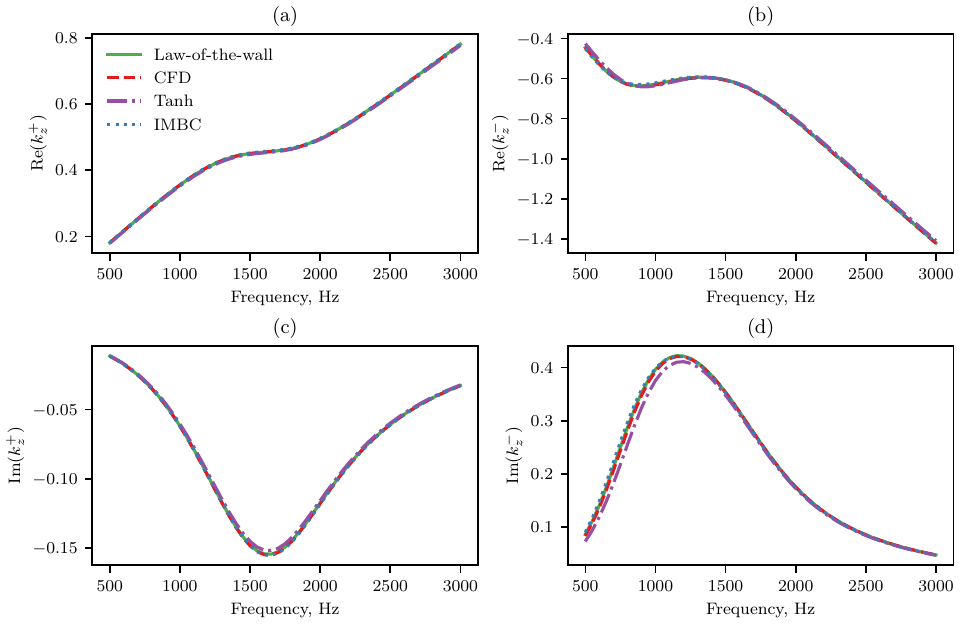}
            \caption{Comparison of the wavenumbers obtained for the reference uniform impedance using different flow profile shapes. $(+)$ denotes downstream propagation, while $(-)$, upstream propagation.}
            \label{fig:2DflowProfileShape}
        \end{figure*}

        First, we analyze the effects of assuming different sheared flow profiles when obtaining the wavenumbers. We compare the solutions obtained using the PBE with the tensorised hyperbolic tangent formulation (Eq.~\eqref{eq:tanh}), the law-of-the-wall (Eq.~\eqref{eq:vanDriest}), and CFD simulations against the solution of the convective Helmholtz equation with the IMBC, all considering the same bulk Mach number, $M$. The resulting wavenumbers are shown in Fig.~\ref{fig:2DflowProfileShape}.
         As a consistency check, wavenumbers generated for a prescribed mean-flow profile and impedance were used as input to the eduction procedure, which is described later in the manuscript. Under identical assumptions, the original impedance was recovered within numerical accuracy. It is observed that, for a given propagation direction, the axial wavenumbers exhibit limited sensitivity to the choice of mean flow profile. This suggests that the details of the shear profile alone are unlikely to explain the discrepancies observed between upstream and downstream impedance eduction, and reinforces the need to consider other factors, such as the effective mean flow Mach number used in the eduction model.
        
        As previously observed for the 2D duct (with a 1D flow profile) in Ref.~\citep{bonomo2026Effect}, good agreement is found between the wavenumbers obtained under the uniform flow assumption and those from the exact solution of the PBE using the law-of-the-wall profile. This suggests that, at least for plane-wavelike modes, simplifying a 3D duct to a 2D or even uniform flow representation may not lead to significant inaccuracies. This contrasts with the findings of Ref.~\citep{roncen2020WavenumberBased}, who considered a less physical hyperbolic tangent flow profile. The wavenumbers obtained considering the CFD flow profile agree well with those from both the IMBC and the tensorised law-of-the-wall profile, suggesting that the latter is a good representation for the duct geometry considered. Therefore, for the sake of conciseness, we adopt the tensorised law-of-the-wall profile as a representative realistic flow profile in the remainder of this work.

    \subsection{On the average Mach number definition}
        
\begin{figure}[htb!]
            \centering
            \includegraphics[width=\linewidth]{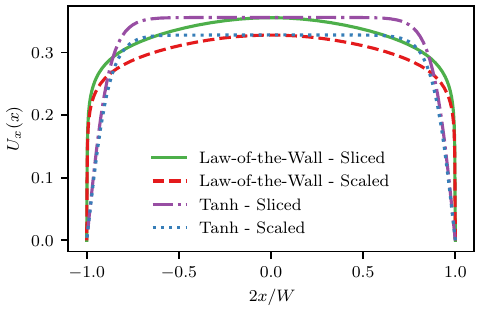}
            \caption{Comparison of the flow profiles considered on the analysis of the average flow profile effect when simplifying to a 2D duct.}
            \label{fig:slicedVsScaled}
        \end{figure}

        Next, we propose an analysis similar to those of Refs.~\cite{roncen2020WavenumberBased} and \cite{jingInvestigationStraightforwardImpedance2015} to investigate the impact of using the ``correct'' bulk Mach number when applying the uniform flow hypothesis or simplifying the geometry to a 2D duct. The key difference from the former lies in our inclusion of upstream-propagating waves, identified as the critical case; from the latter, our focus is on the influence of different flow profile shapes. In what follows, we compare the wavenumbers obtained in the full 3D duct with those from a 2D simplification, using two approaches: in the first (``sliced''), the midspan velocity profile is directly used, while in the second (``scaled''), this same profile is rescaled to match the bulk Mach number, $M$. Hence, the sliced profile has an average Mach number of $M_{1\mathrm{D}}$, while the scaled profile has an average Mach number of $M$ by construction. Since good agreement was previously observed between the CFD velocity profile and the law-of-the-wall representation, and because a consistent 2D reduction of the CFD profile would require a dedicated 2D duct simulation, the CFD-derived profile is omitted here for the sake of brevity. We also compare the results obtained from the IMBC when considering a uniform flow of either $M_{1\mathrm{D}}$ or $M$. The four flow profiles considered are shown in Fig.~\ref{fig:slicedVsScaled}.

        \begin{figure*}[htb!]
            \centering
            \includegraphics[width = 6.4in]{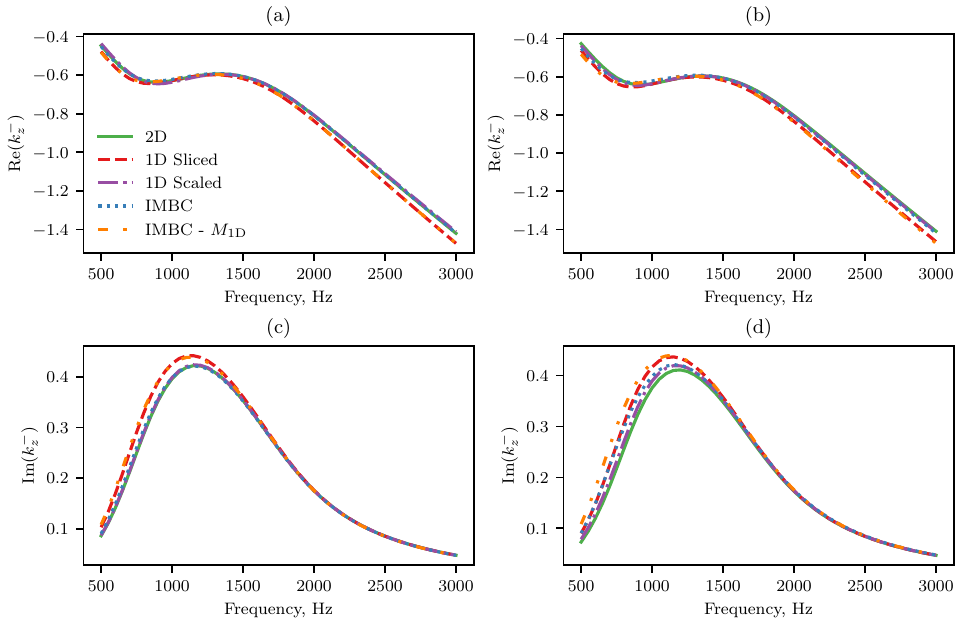}
            \caption{Comparison of wavenumbers for upstream-propagating waves obtained for the reference impedance, using the fully 2D flow profile, simplified 1D profiles with and without scaling to the bulk Mach number, and the IMBC. (a,c) law-of-the-wall; (b,d) hyperbolic tangent profile. }
            \label{fig:2dVs1dVs0d}
        \end{figure*}%
        Fig.~\ref{fig:2dVs1dVs0d} shows the wavenumbers for upstream-propagating waves and the reference impedance, considering both the tensorised law-of-the-wall and hyperbolic tangent profiles. Good agreement is observed between the full 3D solution and the 2D simplification when the 1D profile is scaled to match the bulk Mach number $M$. Moreover, the wavenumbers obtained using the IMBC with uniform flow match those from the scaled case, provided the same average Mach number is used. In contrast, the results for the hyperbolic tangent profile follow the conclusions of Ref.~\cite{roncen2020WavenumberBased}: simplifying from a 3D duct to a 2D geometry yields different wavenumbers for the same impedance, even when the flow profile is scaled. The discrepancy is even greater with the IMBC.

        \begin{figure*}
            \centering
            \includegraphics[width = 6.4in]{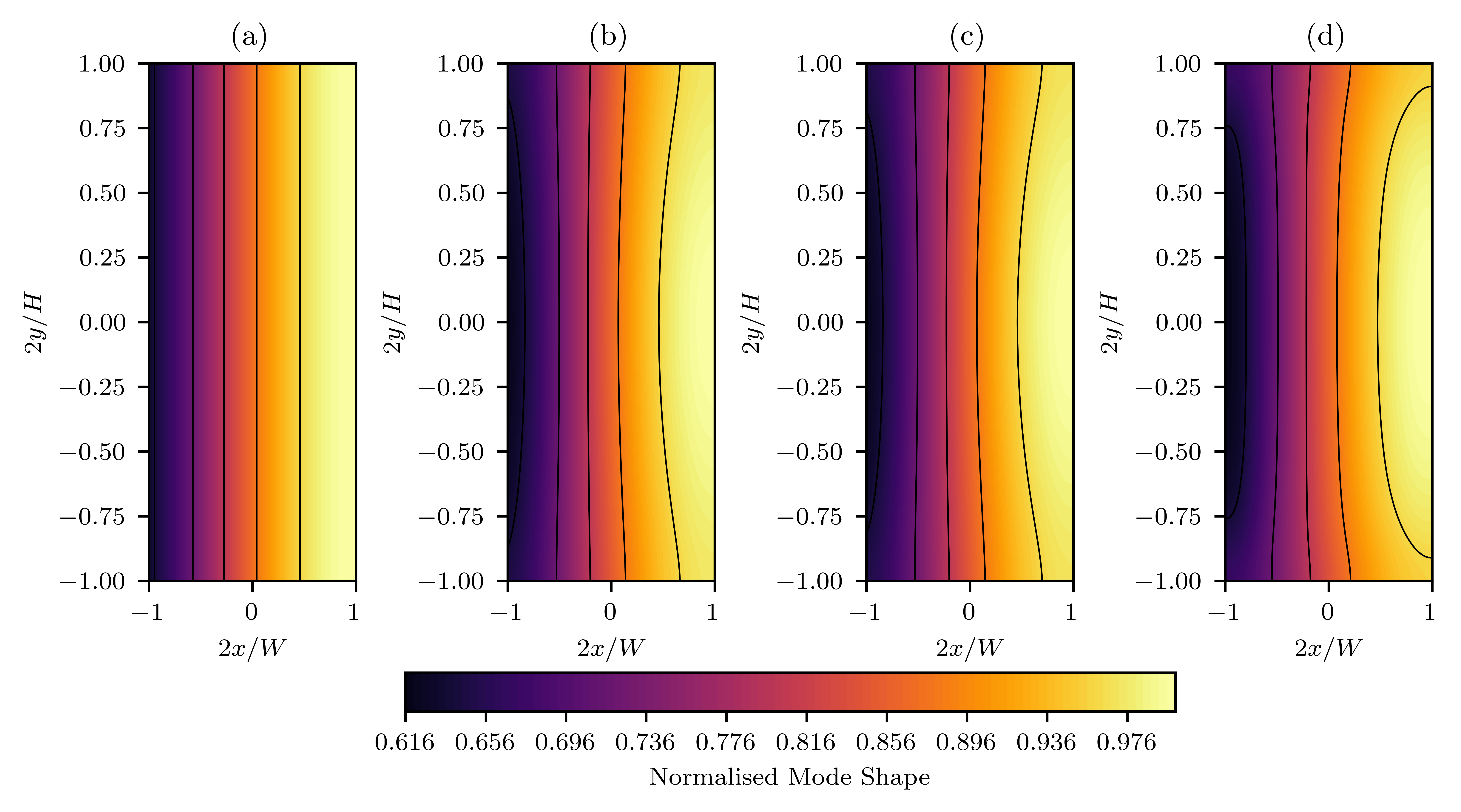}
            \caption{Least attenuated modes obtained for upstream propagation with the reference impedance at \SI{3000}{\hertz}. (a) CHE+IMBC; (b) law-of-the-wall; (c) CFD, and; (d) hyperbolic tangent. Color shading and contour lines represent the same normalized mode shape; contour lines are superimposed only to highlight spatial variations.}
            \label{fig:modeShapes_flowShape}
        \end{figure*}%
         Fig.~\ref{fig:modeShapes_flowShape} shows the normalized acoustic pressure modal shape amplitudes obtained for the reference impedance and the different flow profiles at \SI{3000}{\hertz}. One can observe that the inclusion of a non-uniform axial velocity profile introduces a vertical gradient in the modal shapes. This effect is due to the axial velocity variation across the cross-section and becomes most evident for the hyperbolic tangent profile (Fig.~\ref{fig:modeShapes_flowShape}d), which exhibits a stronger velocity shear near the upper and lower walls compared to the duct centreline.

        The partial conclusions of this study so far can be summarized as follows: (I) The IMBC provides a good approximation for the acoustic modes and wavenumbers in a typical impedance eduction range; and (II) the distortions caused by assuming non-realistic flow profiles may lead to distortions of the acoustic modes and errors in the wavenumbers. 
    
    \subsection{Simulated impedance eduction} \label{sec:eductionShape}
    
        We now replicate another analysis from Ref.~\citep{roncen2020WavenumberBased}, namely the numerical experiment involving synthesized wavenumbers, following a procedure that is more consistent with the framework developed and preliminary results found in this work. In their study, the wavenumbers obtained for the 3D duct were used in an iterative impedance eduction routine, which was then solved assuming a 2D duct geometry. However, only the sliced flow profile was considered, and no reference was made to the traditional eduction using the IMBC as an alternative, on top of considering only the simplified profile given by the hyperbolic tangent formulation.
    
        In what follows, we apply the same procedure, but extend the analysis to include both the sliced and scaled velocity profiles for the 2D duct case, as well as the IMBC formulation using both $M$ and $M_{\mathrm{1D}}$ as input parameters. This numerical experiment mimics practical eduction setups in which the source location relative to the liner determines whether upstream- or downstream-propagating modes are identified and used in the eduction process. The numerical impedance eduction routine can be summarized as follows:
        
        \begin{enumerate}
            \item The PBE eigenvalue solver is used to compute the axial wavenumbers in the lined 3D duct with sheared grazing flow. A pair of axial wavenumbers is obtained, $k^\pm_{z, \mathrm{2D}}$, corresponding to the least attenuated modes. This step mimics a real-world impedance eduction approach where the wavenumbers are extracted from equally spaced acoustic pressure measurements using Prony-like algorithms~\cite{roncen2020WavenumberBased}. In summary, this is the only step that differs our \textit{in silico} experiment from a real world experiment.
            
            \item The following step depends on whether one intends to solve considering a 1D flow profile or assume a uniform flow with IMBC:
            \begin{itemize}
                \item For the 1D case, the educed impedance is obtained by iteratively minimizing the cost function
                \begin{equation}
                    \mathrm{cost~fun} = \left\| k^\pm_{z, \mathrm{2D}} - k^\pm_{z, \mathrm{1D}} \right\|, \label{eq:costFun}
                \end{equation}
                which is solved independently for the upstream and downstream propagation cases;
                
                \item For the uniform flow case with the IMBC, the classical straightforward impedance eduction routine is used, as described in Ref.~\cite{jing2008straightforward}.
            \end{itemize}
        \end{enumerate}

        It is important to emphasize that, in this numerical experiment, the axial wavenumbers computed from the 3D duct with a fully two-dimensional sheared mean flow are treated as reference data, representing an \textit{in silico} impedance eduction experiment. The objective is to assess how the educed impedance is affected when this configuration is simplified to a two-dimensional duct model with a one-dimensional mean-flow profile. Accordingly, the impedance eduction in Eq.~\mbox{\eqref{eq:costFun}} is formulated as the minimization of the discrepancy between the reference axial wavenumbers obtained from the 3D configuration and those predicted by the reduced-order 2D model with a sliced or scaled velocity profile.
    
\begin{figure*}[t]
            \centering
            \includegraphics[width=6.4in]{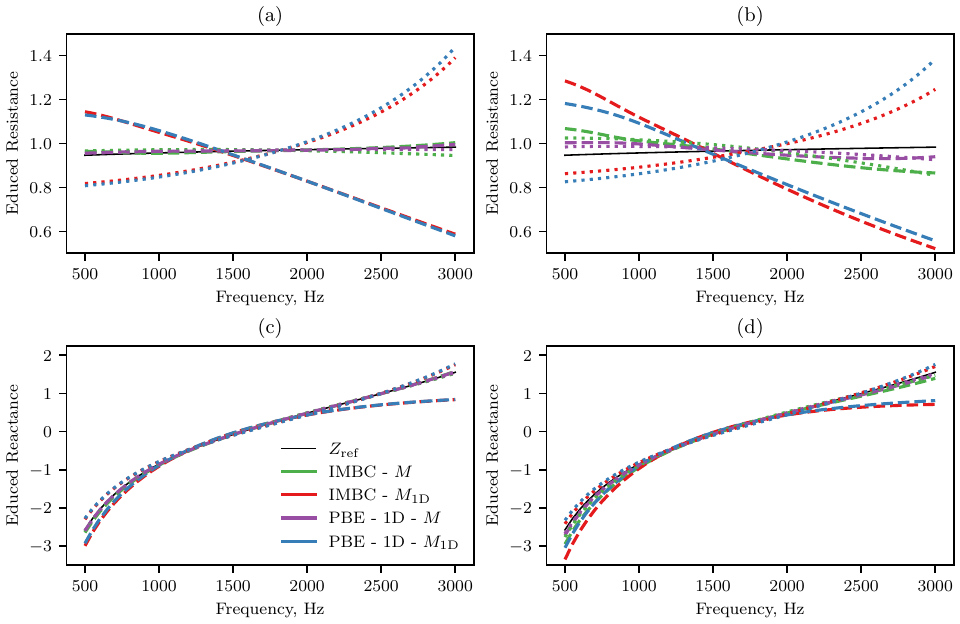}
            \caption{Educed impedances using the proposed numerical experiment, with wavenumbers obtained from the fully 2D (a,c) law-of-the-wall and (b,d) hyperbolic tangent flow profiles. Dashed lines correspond to impedance educed from upstream-propagating modes (downstream source), while dotted lines correspond to downstream-propagating modes (upstream source).}
            \label{fig:eduction1D_machEffect}
        \end{figure*}

        We consider both the tensorised law-of-the-wall and hyperbolic tangent velocity profiles for data generation in this numerical experiment. Fig.~\mbox{\ref{fig:eduction1D_machEffect}} summarizes the outcomes of the \textit{in silico} eduction experiment for different flow representations and modeling assumptions. The figure compares the impedances educed from upstream- and downstream-propagating modes under varying levels of simplification, ranging from reduced-order 1D flow profiles to uniform-flow assumptions using the Ingard--Myers boundary condition. The purpose of this comparison is to assess whether the discrepancies between upstream and downstream eduction can be attributed to these modeling simplifications when the underlying reference data are generated from a realistic two-dimensional mean-flow profile.
        The analysis confirms key trends. First, consistent with Ref.~\cite{roncen2020WavenumberBased}, the reduction to a 2D duct induces the classical mismatch between upstream and downstream propagation observed experimentally when the mean Mach number of the 1D profile, $M_{\mathrm{1D}}$, differs from the bulk Mach number of the 3D duct, $M$. This behavior holds for both velocity profiles considered. This observation is consistent with the NASA impedance eduction protocol \mbox{\cite{nark2018assessment}}, in which a bulk Mach number based on the duct mass flow rate is employed. The present results support the importance of using a consistent effective Mach number in the eduction model to ensure reliable impedance estimation. Second, when the 1D flow profile is scaled to match the original bulk Mach number, this discrepancy vanishes. This supports the findings of Ref.~\cite{jingInvestigationStraightforwardImpedance2015}, who argued that averaging across the full cross-section yields the appropriate Mach number for eduction. Finally, impedances obtained via the traditional straightforward method, which assumes uniform flow and the Ingard--Myers boundary condition, remain a reasonable approximation, as long as the correct average Mach number is used, particularly when realistic flow profiles are employed to perform the \textit{in silico} experiments. Importantly, these results indicate that the uniform-flow assumption and the Ingard–Myers boundary condition are not, by themselves, responsible for the upstream/downstream discrepancy, provided that the reference data are generated using a realistic mean-flow model. 
        
        These results further indicate that the observed upstream/downstream discrepancy is strongly linked to inconsistencies in the effective mean flow Mach number used in the eduction model, rather than to the specific choice of shear profile. When the effective Mach number is not preserved between the reference configuration and the reduced-order model, a mismatch between upstream and downstream propagation naturally arises. Conversely, enforcing this consistency leads to a collapse of the educed impedances, regardless of the level of flow profile simplification.


\section{Conclusions} \label{sec:conclu}

    This paper analyses the effects of two-dimensional (2D) sheared flow profiles on impedance eduction in three-dimensional (3D) ducts, by generating data using an \emph{in silico} acoustics model and then educing impedance using a simplified model. In particular, the influence of adopting simplified velocity profile formulations or geometric simplifications—such as reducing the domain to 2D or assuming uniform flow with the Ingard--Myers Boundary Condition (IMBC)—was investigated.  It was found that simplifying the model used in the eduction process from 3D duct acoustics to a 2D or uniform flow domain does not account for the directional discrepancies in educed impedance observed experimentally, provided that data was generated from \emph{in silico} experiments that used a realistic boundary layer representation. This contrasts with the findings of Ref.~\citep{roncen2020WavenumberBased}, who used a less realistic hyperbolic tangent profile when generating their data. These findings show that the upstream/downstream discrepancies seen in experimental impedance eduction are not an inherent consequence of uniform-flow modeling or the Ingard–Myers boundary condition.
    
    In addition, wavenumbers obtained by solving the Pridmore--Brown equation (PBE) for different sheared flow profiles were compared to those from the Convected Helmholtz Equation (CHE) with IMBC. The results indicate that the IMBC offers a good approximation when compared to solutions based on realistic profiles, such as the universal law-of-the-wall or those derived from CFD simulations, even though these profiles have significant sheared regions across much of the duct width. Conversely, the widely used hyperbolic tangent profile yielded different wavenumbers, even through the boundary layer thickness was matched to give the same boundary layer shape parameters. Additionally, it was shown that when reducing the domain to 2D, the average Mach number of the 1D profile must be scaled to match the bulk Mach number of the full 2D profile.
    
    A key outcome of this study is that the consistency of the effective mean flow Mach number between the reference configuration and the eduction model is a primary factor in reducing the upstream/downstream discrepancy reported in the literature. This result suggests that previously observed differences attributed to shear flow effects may instead arise from inconsistencies in the eduction setup, particularly in the definition of the mean flow used in the propagation model. It is important to note that viscous effects were neglected in this work, and their influence under the same conditions remains to be investigated in future studies. Finally, the conclusions drawn here are based on two simplifying assumptions. First, a single duct geometry was considered; therefore, a parametric investigation of the effect of the duct cross-section is left for future work. Second, the mean flow was assumed to be fully developed and invariant along the duct axis. However, recent findings \mbox{\cite{paduano_impact_2025}} suggest that the axial development of the boundary layer may introduce local effects on the liner impedance.


\hypertarget{ap:utas}{\section*{Appendix A: The Goodrich (or UTAS) semi-empirical model}}

    The Goodrich liner impedance model considered in this work is as given by \citet{yu2008validation}. In this model, the liner impedance is
    \begin{equation}
        Z = Z_{\text{of}} + S_\text{r} U_0 + R_{\text{cm}} + \iu \left( S_\text{m}U_0 - \cot{(kh)} \right), \label{eq:modelBasic}
    \end{equation}
    where $Z_{\text{of}}$ is the perforate plate impedance, $S_\text{r}$ is the non-linear resistance slope, $U_0$ is the root-mean-squared acoustic particle velocity, which is obtained through an iterative process, where the SPL is used as an input\footnote{In this work, non-linear term due to the SPL was neglected.}, $R_{\text{cm}}$ is the normalized grazing flow induced acoustic resistance, $S_\text{m}$ is the non-linear mass reactance, and $h$ is the liner cavity depth.
    
    The impedance $Z_{\text{of}}$ is given by
    \begin{equation}
        Z_{\text{of}} = \iu \omega \dfrac{(t + \varepsilon d)}{c_0 \sigma F\left(\frac{k_s d}{2}\right)},
    \end{equation}
    where $t$ is the facesheet thickness, $d$ is the perforate plate hole diameter, $\sigma$ is the percentage of open area, $F(k_s d/2)$  is the cross-section averaged hole velocity profile from Crandall's solution~\citep{crandall1926theory} and 
    \begin{equation}
        \varepsilon d = \dfrac{d(1-0.7\sqrt{\sigma})}{1+305M^3}
    \end{equation}
    is the effective mass end correction~\citep{rice1971}. The cross-section-averaged hole velocity profile is defined as
    \begin{equation}
        F\left(\frac{k_s d}{2}\right) = 1 - \dfrac{2J_1\left(\frac{k_s d}{2}\right)}{\frac{k_s d}{2} J_0 \left(\frac{k_s d}{2}\right)},
    \end{equation}
    where $J_0$ and $J_1$ are zero- and first-order Bessel functions,
    \begin{equation}
        k_s^2 = -\iu \dfrac{\omega \rho_0}{\mu}
    \end{equation}
    is the wavenumber of a viscous Stokes wave and $\mu$ is the air viscosity.
    
    The non-linear resistance slope is given by
    \begin{equation}
        S_\text{r} = 1.336541\left( \dfrac{1-\sigma^2}{2c_0C_d^2\sigma^2} \right),
    \end{equation}
    where $C_d$ is the discharge coefficient, which for $t/d\leq1$ is given by
    \begin{equation}
        C_d = 0.80695\sqrt{\dfrac{\sigma^{0.1}}{\exp{\left(\frac{-0.5072t}{d}\right)}}}.
    \end{equation}
    
    The normalized acoustic resistance under grazing flow is given by the Rice--Heidelberg derivation \citep{heidelberg1980Experimental}, such that
    \begin{equation}
        R_{\text{cm}} = \dfrac{M}{\sigma\left( 2+1.256 \dfrac{\delta^*}{d} \right)},
    \end{equation}
    where $\delta^*$ is the flow profile boundary layer displacement thickness.
    Finally, the non-linear mass reactance is given by
    \begin{equation}
        S_\text{m} = -0.0000207\dfrac{k}{\sigma^2}.
    \end{equation}

\section*{Acknowledgments}
    On behalf of  L.A.~Bonomo and J.A.~Cordioli, this research was partially funded by CNPq (National Council for Scientific and Technological Development).
    L.A.~Bonomo acknowledges that this study was financed in part by the Coordenação de Aperfeiçoamento de Pessoal de Nível Superior – Brasil (CAPES), Finance Code 001.
    E.J.~Brambley gratefully acknowledges the support of the UK Engineering and Physical Sciences Research Council (EPSRC grant EP/V002929/1). The work of A.~Paduano and F.~Avallone is funded by the European Union (ERC, LINING, 101075903). Views and opinions expressed are however those of the author(s) only and do not necessarily reflect those of the European Union or the European Research Council. Neither the European Union nor the granting authority can be held responsible for them.

\raggedright
\bibliography{sample}

@string(JFM="J.~Fluid Mech.")

@string(JSV="J.~Sound Vib.")

@string(JASA="J.~Acoust. Soc. Am.")

@string(JAS="J.~Aeronaut. Sci.")

@string(AIAAJ="{AIAA}~J.")

@string(AA="Appl. Acoust.")

@string(IJA="{Int. J.~Aeroacoust.}")

@article{nakayama1983Calculation,
  title = {Calculation of Fully Developed Turbulent Flows in Ducts of Arbitrary Cross-Section},
  author = {Nakayama, A. and Chow, W. L. and Sharma, D.},
  year = {1983},
  journal = {Journal of Fluid Mechanics},
  volume = {128},
  pages = {199--217},
  issn = {1469-7645, 0022-1120},
  doi = {10.1017/S0022112083000440}
}

@article{guess1975Calculation,
  title = {Calculation of Perforated Plate Liner Parameters from Specified Acoustic Resistance and Reactance},
  author = {Guess, A. W.},
  year = {1975},
  month = may,
  journal = JSV,
  volume = {40},
  number = {1},
  pages = {119--137},
  publisher = {{Academic Press}},
  issn = {0022-460X},
  doi = {10.1016/S0022-460X(75)80234-3}
}

@techreport{jones2002effects,
  title = {Effects of Liner Geometry on Acoustic Impedance},
  booktitle = {8th {AIAA}/{CEAS} Aeroacoustics Conference},
  institution = {8th {AIAA}/{CEAS} Aeroacoustics Conference},
  author = {Jones, Michael and Tracy, Maureen and Watson, Willie and Parrott, Tony},
  year = {2002},
  publisher = {{AIAA}},
  address = {{Breckenridge, Colorado}},
  doi = {10.2514/6.2002-2446},
  type = {{AIAA} Paper 2002-2446}
}

@techreport{murray2012development,
  title = {Development of a Single Degree of Freedom Perforate Impedance Model under Grazing Flow and High {SPL}},
  booktitle = {18th {AIAA}/{CEAS} Aeroacoustics Conference (33rd {AIAA} Aeroacoustics Conference)},
  institution = {18th {AIAA}/{CEAS} Aeroacoustics Conference (33rd {AIAA} Aeroacoustics Conference)},
  author = {Murray, Paul and Astley, R. Jeremy},
  year = {2012},
  month = jun,
  publisher = {{AIAA}},
  address = {{Colorado Springs, Colorado}},
  doi = {10.2514/6.2012-2294},
  isbn = {978-1-60086-932-7},
  type = {{AIAA} Paper 2012-2294}
}

@techreport{kooi1981Experimental,
  title = {An Experimental Study of the Acoustic Impedance of {H}elmholtz Resonator Arrays under a Turbulent Boundary Layer},
  booktitle = {7th {AIAA} Aeroacoustics Conference},
  institution = {7th {AIAA} Aeroacoustics Conference},
  author = {Kooi, J. and Sarin, S.},
  year = {1981},
  month = oct,
  publisher = {{AIAA}},
  address = {{Palo Alto, CA}},
  doi = {10.2514/6.1981-1998},
  type = {{AIAA} Paper 81-1998},
}

@article{watson1999validation,
  title = {Validation of an Impedance Eduction Method in Flow},
  author = {Watson, W and Jones, M and Parrott, T},
  year = {1999},
  journal = AIAAJ,
  volume = {37},
  number = {7},
  pages = {818--824},
  doi = {10.2514/2.7529}
}

@misc{paduano_impact_2025,
	title = {On the impact of the turbulent grazing flow development on the acoustic response of an acoustic liner},
	doi = {10.48550/arXiv.2507.22714},
	publisher = {arXiv},
	author = {Paduano, Angelo and Scarano, Francesco and Cordioli, Julio and Casalino, Damiano and Avallone, Francesco},
	month = jul,
	year = {2025},
	note = {{Preprint} available at arXiv:2507.22714 [physics]}
}

@article{jing2008straightforward,
  title = {A Straightforward Method for Wall Impedance Eduction in a Flow Duct},
  author = {Jing, Xiaodong and Peng, Sen and Sun, Xiaofeng},
  year = {2008},
  journal = JASA,
  volume = {124},
  number = {1},
  pages = {227--234},
  issn = {0001-4966},
  doi = {10.1121/1.2932256}
}

@article{elnady2009validation,
  title = {Validation of an Inverse Semi-Analytical Technique to Educe Liner Impedance},
  author = {Elnady, T. and Bod{\'e}n, H. and Elhadidi, B.},
  year = {2009},
  journal = AIAAJ,
  volume = {47},
  number = {12},
  pages = {2836--2844},
  issn = {0001-1452},
  doi = {10.2514/1.41647}
}

@article{ferrante2016back,
  title = {Back-to-Back Comparison of Impedance Measurement Techniques Applied to the Characterization of Aero-Engine Nacelle Acoustic Liners},
  author = {Ferrante, Piergiorgio and De Roeck, Wim and Desmet, Wim and Magnino, Nicola},
  year = {2016},
  journal = AA,
  volume = {105},
  pages = {129--142},
  issn = {1872910X},
  doi = {10.1016/j.apacoust.2015.12.004}
}

@article{bonomoComparisonSituImpedance2024,
  title = {A Comparison of in Situ and Impedance Eduction Experimental Techniques for Acoustic Liners with Grazing Flow and High Sound Pressure Level},
  author = {Bonomo, Lucas A and Quintino, Nicolas T and Spillere, Andr{\'e} M N and Murray, Paul B and Cordioli, Julio A},
  year = {2024},
  month = jan,
  journal = IJA,
  pages = {60-83},
volume = {23(1-2)},
  publisher = {{SAGE Publications}},
  issn = {1475-472X},
  doi = {10.1177/1475472X231225629}
}

@article{renou2011Failure,
  title = {Failure of the {I}ngard--{M}yers Boundary Condition for a Lined Duct: {A}n Experimental Investigation},
  author = {Renou, Yga{\"a}l and Aur{\'e}gan, Yves},
  year = {2011},
  month = jul,
  journal = JASA,
  volume = {130},
  number = {1},
  pages = {52},
  publisher = {{Acoustical Society of AmericaASA}},
  issn = {0001-4966},
  doi = {10.1121/1.3586789}
}

@techreport{boden2016Effect,
  title = {On the Effect of Flow Direction on Impedance Eduction Results},
  booktitle = {22nd {AIAA}/{CEAS} Aeroacoustics Conference},
  institution = {22nd {AIAA}/{CEAS} Aeroacoustics Conference},
  author = {Bod{\'e}n, Hans and Zhou, Lin and Cordioli, Julio A. and Medeiros, Augusto A. and Spillere, Andr{\'e} M.N.},
  year = {2016},
  month = jun,
  publisher = {{AIAA}},
  address = {{Lyon, France}},
  doi = {10.2514/6.2016-2727},
  type = {{AIAA} Paper 2016-2727},
  isbn = {978-1-62410-386-5}
}

@techreport{roncen2019influence,
  title = {Influence of Source Propagation Direction and Shear Flow Profile in Impedance Eduction of Acoustic Liners},
  booktitle = {25th {AIAA}/{CEAS} Aeroacoustics Conference},
  institution = {25th {AIAA}/{CEAS} Aeroacoustics Conference},
  author = {Roncen, Remi and Piot, Estelle and Mery, Fabien and Simon, Frank and Jones, Michael G. and Nark, Douglas M.},
  year = {2019},
  publisher = {{AIAA}},
  address = {{Delft, The Netherlands}},
  doi = {10.2514/6.2019-2469},
  type = {{AIAA} Paper 2019-2469}
}

@techreport{boden2017Comparison,
  title = {Comparison of the Effect of Flow Direction on Liner Impedance Using Different Measurement Methods},
  booktitle = {23rd {AIAA}/{CEAS} Aeroacoustics Conference},
  institution = {23rd {AIAA}/{CEAS} Aeroacoustics Conference},
  author = {Bod{\'e}n, Hans and Cordioli, Julio A. and Spillere, Andr{\'e} M.N. and Serrano, Pablo G.},
  year = {2017},
  month = jun,
  publisher = {{AIAA}},
  address = {{Denver, Colorado}},
  doi = {10.2514/6.2017-3184},
  isbn = {978-1-62410-504-3},
  type = {{AIAA} Paper 2017-3184}
}

@techreport{nark2018assessment,
  title = {Assessment of Axial Wave Number and Mean Flow Uncertainty on Acoustic Liner Impedance Eduction},
  booktitle = {2018 {AIAA}/{CEAS} Aeroacoustics Conference},
  institution = {2018 {AIAA}/{CEAS} Aeroacoustics Conference},
  author = {Nark, Douglas M. and Jones, Michael G. and Piot, Estelle},
  year = {2018},
  month = jun,
  publisher = {{AIAA}},
  address = {{Reston, Virginia}},
  doi = {10.2514/6.2018-3444},
  type = {{AIAA} Paper 2018-3444},
  isbn = {978-1-62410-560-9}
}

@article{roncen2020WavenumberBased,
  title = {Wavenumber-{{Based Impedance Eduction}} with a {{Shear Grazing Flow}}},
  author = {Roncen, R. and Piot, E. and M{\'e}ry, F. and Simon, F. and Jones, M. G. and Nark, D. M.},
  year = {2020},
  journal = AIAAJ,
  volume = {58},
  number = {7},
  pages = {3040--3050},
  publisher = {{American Institute of Aeronautics and Astronautics}},
  issn = {0001-1452},
  doi = {10.2514/1.J059100}
}

@article{bonomo2026Effect,
  title = {Effects of the Sheared Flow Velocity Profile on Impedance Eduction in a {{2D}} Duct},
  author = {Bonomo, Lucas Araujo and Brambley, Edward James and Cordioli, Julio Apolinario},
  year = 2026,
  month = jan,
  journal = {Acta Acustica},
  volume = {10},
  pages = {6},
  publisher = {EDP Sciences},
  issn = {2681-4617},
  doi = {10.1051/aacus/2026005},
  urldate = {2026-01-26},
  copyright = {\copyright{} 2026, The Author(s), Published by EDP Sciences},
  langid = {english}
}

@article{nayfeh1974Effect,
  title = {Effect of Mean-Velocity Profile Shapes on Sound Transmission through Two-Dimensional Ducts},
  author = {Nayfeh, A. H. and Kaiser, J. E. and Shaker, B. S.},
  year = {1974},
  month = jun,
  journal = JSV,
  volume = {34},
  number = {3},
  pages = {413--423},
  issn = {0022-460X},
  doi = {10.1016/S0022-460X(74)80320-2}
}

@article{pridmorebrown1958sound,
  title = {Sound Propagation in a Fluid Flowing through an Attenuating Duct},
  author = {Pridmore-Brown, D. C.},
  year = {1958},
  month = aug,
  journal = JFM,
  volume = {4},
  number = {04},
  pages = {393--406},
  publisher = {{Cambridge University Press}},
  issn = {0022-1120},
  doi = {10.1017/S0022112058000537}
}

@article{ingard1959influence,
  title = {Influence of Fluid Motion Past a Plane Boundary on Sound Reflection, Absorption, and Transmission},
  author = {Ingard, Uno},
  year = {1959},
  journal = JASA,
  volume = {31},
  number = {7},
  pages = {1035--1036},
  issn = {0001-4966},
  doi = {10.1121/1.1907805}
}

@article{myers1980acoustic,
  title = {On the Acoustic Boundary Condition in the Presence of Flow},
  author = {Myers, M K},
  year = {1980},
  journal = JSV,
  volume = {71},
  pages = {429--434},
  issn = {0022460X},
  doi = {10.1016/0022-460X(80)90424-1}
}

@article{boyer2011Theoretical,
  title = {Theoretical Investigation of Hydrodynamic Surface Mode in a Lined Duct with Sheared Flow and Comparison with Experiment},
  author = {Boyer, Germain and Piot, Estelle and Brazier, Jean-Philippe},
  year = {2011},
  month = apr,
  journal = JSV,
  volume = {330},
  number = {8},
  pages = {1793--1809},
  issn = {0022-460X},
  doi = {10.1016/j.jsv.2010.10.035}
}

@book{boyd2001Chebyshev,
  title = {{C}hebyshev and {F}ourier Spectral Methods},
  author = {Boyd, J.P.},
  year = {2001},
  edition = {2},
  publisher = {Dover (Courier Corporation)},
address = {Garden City, NY, USA},
  isbn = {978-0486411835}
}

@techreport{yu2008Validation,
  title = {Validation of {G}oodrich Perforate Liner Impedance Model Using {NASA} {L}angley Test Data},
  booktitle = {14th {AIAA}/{CEAS} Aeroacoustics Conference (29th {AIAA} Aeroacoustics Conference)},
  institution = {14th {AIAA}/{CEAS} Aeroacoustics Conference (29th {AIAA} Aeroacoustics Conference)},
  author = {Yu, Jia and Ruiz, Marta and Kwan, Hwa Wan},
  year = {2008},
  month = {5},
  address = {{Vancouver, British Columbia, Canada}},
  doi = {10.2514/6.2008-2930},
  isbn = {978-1-56347-939-7},
  type = {{AIAA} Paper 2008-2930}
}

@techreport{pereira2023Validation,
  title = {Validation of High-Fidelity Numerical Simulations of Acoustic Liners Under Grazing Flow},
  booktitle = {{AIAA} AVIATION},
  institution = {{AIAA} AVIATION},
  author = {Pereira, Lucas and Bonomo, Lucas and Quintino, Nicolas and {da Silva}, Andrey and Cordioli, Julio and Avallone, Francesco},
  year = {2023},
  address = {{San Diego, USA}},
  doi = {10.2514/6.2023-3503},
  type = {{AIAA} Paper 2023-3503}
}

@techreport{rienstra2008spatial,
  title={Spatial instability of boundary layer along impedance wall},
  author={Rienstra, Sjoerd and Vilenski, Gregory},
  booktitle={14th {AIAA}/{CEAS} Aeroacoustics Conference},
  institution={14th {AIAA}/{CEAS} Aeroacoustics Conference},
  year={2008},
  doi={10.2514/6.2008-2932},
  type={{AIAA} Paper 2008-2932}
}

@article{van1956turbulent,
  title={On turbulent flow near a wall},
  author={Van Driest, Edward R},
  journal=JAS,
  volume={23},
  number={11},
  pages={1007--1011},
  year={1956},
  doi = {10.2514/8.3713},
}

@article{jingInvestigationStraightforwardImpedance2015,
  title = {Investigation of Straightforward Impedance Eduction in the Presence of Shear Flow},
  author = {Jing, Xiaodong and Peng, Sen and Wang, Lixun and Sun, Xiaofeng},
  year = {2015},
  month = jan,
  journal = JSV,
  volume = {335},
  pages = {89--104},
  issn = {0022-460X},
  doi = {10.1016/j.jsv.2014.08.031}
}

@inproceedings{menter2003ten,
  title={Ten years of industrial experience with the {SST} turbulence model},
  author={Menter, Florian R and Kuntz, Martin and Langtry, Robin},
  booktitle={Turbulence, Heat and Mass Transfer},
  volume={4},
  pages={625--632},
  year={2003},
  note={In Proc. 4th International Symposium on Turbulence, Heat and Mass Transfer, Antalya, Turkey},
  url = {https://www.researchgate.net/publication/228742295_Ten_years_of_industrial_experience_with_the_SST_turbulence_model}
}

@book{crandall1926theory,
  title={Theory of vibrating systems and sound},
  author={Crandall, Irving Bardshar},
  year={1926},
  publisher={Van Nostrand}
}

@techreport{rice1971,
 type = {{NASA} Technical Report},
  title={Acoustic and aerodynamic performance of a 6-foot-diameter fan for turbofan engines. 3 - {P}erformance with noise suppressors},
  author={Rice, E J and Feiler, C E and Acker, L W},
  year={1971},
  number = {NASA-TN–D-6178},
  institution = {{NASA}},
  url={https://ntrs.nasa.gov/citations/19710007779}
}

@techreport{heidelberg1980Experimental,
  type = {{NASA} Technical Report},
  title = {Experimental Evaluation of a Spinning-Mode Acoustic-Treatment Design Concept for Aircraft Inlets},
  author = {Heidelberg, L.J. and Rice, E.J. and Homyak, L.},
  year = {1980},
  month = apr,
  number = {NASA-TP-1613},
  institution = {{NASA}},
  url={https://ntrs.nasa.gov/citations/19800012838}
}

@article{weng2018Flow,
  title = {Flow and {{Viscous Effects}} on {{Impedance Eduction}}},
  author = {Weng, Chenyang and Schulz, Anita and Ronneberger, Dirk and Enghardt, Lars and Bake, Friedrich},
  year = {2018},
  month = mar,
  journal = AIAAJ,
  volume = {56},
  number = {3},
  pages = {1118--1132},
  publisher = {{American Institute of Aeronautics and Astronautics}},
  issn = {0001-1452},
  doi = {10.2514/1.J055838}
}

@article{brambley2009fundamental,
  title = {Fundamental Problems with the Model of Uniform Flow over Acoustic Linings},
  author = {Brambley, Edward James},
  year = {2009},
  journal = {Journal of Sound and Vibration},
  volume = {322},
  number = {4-5},
  pages = {1026--1037},
  issn = {0022460X},
  doi = {10.1016/j.jsv.2008.11.021}
}

@article{yangShearFlowEffects2024,
  title = {Shear Flow Effects in a {{2D}} Duct: {{Influence}} on Wave Propagation and Direct Impedance Eduction},
  shorttitle = {Shear Flow Effects in a {{2D}} Duct},
  author = {Yang, Jinyue and Humbert, Thomas and Golliard, Joachim and Gabard, Gw{\'e}na{\"e}l},
  year = {2024},
  month = apr,
  journal = {Journal of Sound and Vibration},
  volume = {576},
  pages = {118296},
  issn = {0022-460X},
  doi = {10.1016/j.jsv.2024.118296},
  urldate = {2024-05-15}}

@article{spillere2020Experimentally,
  title = {Experimentally Testing Impedance Boundary Conditions for Acoustic Liners with Flow: {{Beyond}} Upstream and Downstream},
  author = {Spillere, Andr{\'e} Mateus Netto and Bonomo, Lucas Araujo and Cordioli, J{\'u}lio Apolin{\'a}rio and Brambley, Edward James},
  year = {2020},
  month = dec,
  journal = {Journal of Sound and Vibration},
  volume = {489},
  pages = {115676},
  publisher = {{Academic Press}},
  issn = {0022-460X},
  doi = {10.1016/J.JSV.2020.115676}
}

@article{zhang2016Numerical,
  title = {Numerical Investigation of a Honeycomb Liner Grazed by Laminar and Turbulent Boundary Layers},
  author = {Zhang, Qi and Bodony, Daniel J.},
  year = {2016},
  month = apr,
  journal = {Journal of Fluid Mechanics},
  volume = {792},
  pages = {936--980},
  publisher = {{Cambridge University Press}},
  issn = {0022-1120, 1469-7645},
  doi = {10.1017/jfm.2016.79},
  langid = {english}
}

@article{rashidi20253D,
  title = {{{3D Multimodal}} Inverse Method for Liner Impedance Eduction},
  author = {Rashidi, Hamid and Golliard, Joachim and Humbert, Thomas},
  year = 2025,
  month = jan,
  journal = {Journal of Sound and Vibration},
  pages = {118954},
  issn = {0022-460X},
  doi = {10.1016/j.jsv.2025.118954},
  urldate = {2025-02-10}}

@article{troian2017Broadband,
  title = {Broadband Liner Impedance Eduction for Multimodal Acoustic Propagation in the Presence of a Mean Flow},
  author = {Troian, Renata and Dragna, Didier and Bailly, Christophe and Galland, Marie-Annick},
  date = {2017-03-31},
  journal = {Journal of Sound and Vibration},
  volume = {392},
  pages = {200--216},
  issn = {0022-460X},
  year = {2017},
  doi = {10.1016/j.jsv.2016.10.014}}

@article{zhao2025DeepLearningBased,
  title = {Deep-{{Learning-Based Impedance Eduction Method}} for {{Acoustic Liners}} with {{Flow}}},
  author = {Zhao, Yao and Deng, Yuanyuan and Lu, Huancai and Jiang, Hanbo},
  year = {2025},
  date = {2025-10-13},
  journal = {AIAA Journal},
  pages = {1--15},
  publisher = {{American Institute of Aeronautics and Astronautics}},
  issn = {0001-1452},
  doi = {10.2514/1.J065518}}

@article{quintino2025Comparison,
  title = {Comparison of {{Impedance Eduction Test Rigs}} with {{Different Boundary-Layer Profiles}}},
  author = {Quintino, Nicolas T. and Bonomo, Lucas A. and Cordioli, Julio A. and Jones, Michael G. and Howerton, Brian M. and Nark, Douglas M. and Avallone, Francesco},
  year = 2025,
  month = nov,
  journal = {AIAA Journal},
  volume = {63},
  number = {11},
  pages = {4872--4883},
  publisher = {{American Institute of Aeronautics and Astronautics}},
  issn = {0001-1452},
  doi = {10.2514/1.J065173}}

\end{document}